
\magnification 1200
\hsize 16.2 true cm             
\baselineskip = 7.600truemm        
\def\GC#1{\beta_{#1}}
\def\GCdd{\beta_2^2}
\def\Ztd{\zeta^2(3)}
\def\Qtw{Q_{12}}
\def\r {r}

\def\lr{\ln \r}
\def\lrd{\ln^2 \r}
\def\lrt{\ln^3 \r}
\def\lrq{\ln^4 \r}

\def\LGd{\ln 2}
\def\LGdd{\ln^2 2}
\def\LGdt{\ln^3 2}
\def\LGdq{\ln^4 2}
\def\LGdc{\ln^5 2}
\def\A#1{a_{#1}}
\def\Z#1{\zeta(#1)}
\def\parn{\par\noindent}

\def\PR{{\it Phys.Rev. }}

\def\ref#1{[#1]}
\def\oo{\infty}

\def\Li{{\rm Li}}

\def\fig#1{{\rm fig. #1} }

\def\+{\hskip -2pt + \hskip -2pt}

\def\amu{{a_\mu^{(5)}}}

\def\#{{\hskip -.55 pt}}
\def\riga{\par\noindent
----------------------------------------------------------------------------------------------------------------
\par\noindent}
\def\d${$ \displaystyle }

\rightline{\bf DFUB 94-02}
\rightline{23 March 1994}
\vskip 20truemm
\centerline{\bf Analytical and numerical contributions of some
                tenth-order graphs}
\par
\centerline{\bf containing vacuum polarization insertions }
\par
\centerline{\bf to the muon (g-2) in QED. }
\par
\vskip 20truemm
\centerline{S.Laporta
\footnote {$^{\S}$}{ {E-mail: {\tt laporta@bo.infn.it} }}  }
\vskip 20truemm
\centerline{\it Dipartimento di Fisica, Universit\`a di Bologna,}
\centerline{\it and INFN, Sezione di Bologna,}
\centerline{\it Via Irnerio 46, I-40126 Bologna, Italy}
\vskip 20truemm
\centerline{\bf Abstract. }  \par
The contributions to the $g-2$ of the muon from some
tenth-order (five-loop) graphs
containing one-loop and two-loop vacuum polarization insertions
have been evaluated analytically in QED perturbation theory,
expanding the results in the
ratio of the electron to muon mass $(m_e / m_\mu)$.
Some results contain also terms known only in numerical form.
Our results agree with the renormalization group results
already existing in the literature.
\vfill\eject

In this work we have calculated in analytical form
the contributions
to the muon $g$-$2$ from the five-loop graphs
obtained inserting
one-loop and two-loop vacuum
polarizations on a single photon line of the second-order and
fourth-order vertex graphs;
the considered graphs are shown in figs.1 and 2.
Following the method used in previous calculations of the
contributions of
three-loop and four-loop graphs containing vacuum polarizations [1][2],
we found it convenient to express the results
expanding them in the ratio of the electron and muon masses
\d${(m_e/m_\mu)}$.
Due to the length of the expressions of the coefficients of
the high power terms, we will list them only up to \d$(m_e/m_\mu)$.

The analytical expressions of the contributions to the muon anomaly
of the graphs shown in figs.1 and 2,
accounting for the proper multiplicity factors,
are ($r\equiv m_e/m_\mu$):
$$ \eqalign { \amu[{\rm fig.1(a)}]=
& {8\over81} \lrq
 +{200\over243} \lrt
 +\left( {8\over81} \pi^2 +{634\over243} \right) \lrd
+\left( {16\over27}\Z3 +{100\over243} \pi^2
\right.\cr
&\left.
+{8609\over2187} \right) \lr
+{2\over135} \pi^4 +{100\over81}\Z3 +{317\over729} \pi^2
                     +{64613\over26244}
\cr
&
-\r\left[ {18203\over374220} \pi^4 \right]
          +O(\r^2\lrq)\ ,}  \eqno(1) $$
$$ \eqalign { \amu[{\rm fig.1(b)}]=
 &\left( {32\over81} \pi^2 -{952\over243} \right) \lrt
 +\left(  {8\over81} \pi^2 -{244\over243} \right) \lrd
+\left( {32\over135} \pi^4  -{104\over81} \pi^2
\right.\cr
&\left.
 -{7627\over729}
                              \right) \lr
 +{16\over27} \pi^2\Z3 +{8\over405} \pi^4 -{476\over81}\Z3
      -{2593\over2187} \pi^2 +{64244\over6561}
\cr
&
          +O(\r^2\lrd)\ ,}  \eqno(2) $$
$$ \eqalign { \amu[{\rm fig.1(c)}]=
&\left( {64\over9}\Z3 -{32\over405} \pi^2 -{1886\over243} \right) \lrd
+\left( -{16\over81} \pi^4 +{16\over135}\Z3 +{21532\over6075} \pi^2
\right.\cr
&\left.
       -{57899\over3645} \right) \lr
+{64\over9}\Z5 +{32\over27} \pi^2\Z3
-{106\over6075} \pi^4 +{10732\over2025}\Z3
\cr
&
-{148921\over91125} \pi^2 -{1090561\over109350}
          +O(\r^2\lr) \ ,}  \eqno(3) $$
$$ \eqalign { \amu[{\rm fig.1(d)}]=
&\left(  {16\over135} \pi^4 -{256\over189}\Z3 -{151849\over15309} \right) \lr
 -{16\over9} \pi^2\Z3
 +{124\over8505} \pi^4
\cr
 &
+{92476\over6615}\Z3 +{143\over81} \pi^2 -{46796257\over3214890}
          +O(\r^2) \ ,}  \eqno(4) $$
$$ \eqalign { \amu[{\rm fig.1(e)}]=
&
 -{1\over3} \lrt
 +\left(  {2\over3}\Z3 -{35\over18} \right) \lrd
 +\left( {25\over9}\Z3 -{1\over6} \pi^2 -{413\over108} \right) \lr
\cr
&
+{1\over9} \pi^2\Z3
+{263\over108}\Z3 -{35\over108} \pi^2
 -{439\over162}
\cr
&
+\r\left[
        -{2516\over945} \pi^3\GC2 -{3401\over2268} \pi^5
         -{11\over54} \pi^4 \LGd   -{1549\over360} \pi^2\Z3
\right.\cr
&
\phantom{+\r[}
\left.
         +{12195383\over28576800} \pi^4+{87998\over8505} \pi^3
         +{970237\over43740} \pi^2
\right]
          +O(\r^2\lrq) \ ,}  \eqno(5) $$
$$ \eqalign { \amu[{\rm fig.1(f)}]=
&
   \left( -{2\over3} \pi^2 +{119\over18} \right) \lrd
  +\left(  {4\over3} \pi^2 \Z3 -{119\over9}\Z3
          +{1\over6} \pi^2   -{13\over8}   \right) \lr
\cr
	   &+{1\over9} \pi^2\Z3  -{2\over15} \pi^4
          -{61\over54}\Z3 +{161\over216} \pi^2 +{3661\over648}
            +O(r^2\lrd)\ ,}  \eqno(6) $$
$$ \eqalign { \amu[{\rm fig.1(g)}]=
&
\left( -{4}\Z3 +{2\over45} \pi^2 +{943\over216} \right) \lr
 +{8}\Ztd  -{4\over45} \pi^2\Z3
\cr
&
 +{1\over18} \pi^4
-{3833\over540}\Z3 -{5483\over5400} \pi^2   +{8581\over3240}
            +O(r^2\lr)\ ,}  \eqno(7) $$
$$ \eqalign { \amu[{\rm fig.1(h)}]=
&\left(  -{14\over405} \pi^4 +{128\over9} \A4
        -{16\over27}  \pi^2 \LGdd  +{16\over27} \LGdq
        +{26\over27} \Z3 -{16\over27} \pi^2 \LGd
\right.\cr
&\left.
        -{164\over243} \pi^2  +{673\over81}
	                       \right) \lrd
+\left( -{146\over9} \Z5
      +{256\over9} \A5
      -{76\over27}  \pi^2\Z3
      +{32\over81}  \pi^2 \LGdt
\right.\cr
&\left.
      +{196\over405}  \pi^4 \LGd
      -{32\over135}  \LGdc
      -{8\over27}  \pi^2 \LGdd
      +{16\over3}  \pi^2 \LGd
      -{44\over405}  \pi^4
\right.\cr
&\left.
      +{64\over3} \A4
      +{8\over9}  \LGdq
      -{1213\over81} \Z3
      -{4873\over1458}  \pi^2
      +{33335\over1944}           \right) \lr
      +U_1
      -{463\over27}\Z5
\cr
&
      +{64\over3}\A5
      +{29\over135} \pi^4 \LGd
      -{161\over162} \pi^2\Z3
      +{8\over81} \pi^2 \LGdt
      -{8\over45} \LGdc
\cr
&
      +{301\over7290} \pi^4
      -{1696\over81}\A4
      -{436\over243} \pi^2 \LGdd
      -{212\over243} \LGdq
      -{3211\over324}\Z3
\cr
&
      -{952\over243} \pi^2 \LGd
      +{2635\over648} \pi^2
      +{31603\over1944}
            +O(\r^2\lr) \ ,}  \eqno(8) $$
$$ \eqalign { \amu[{\rm fig.1(i)}]=
&\left({175\over9}\Z5          +{140\over27} \pi^2\Z3
      -{10\over9} \pi^4 \LGd   +{14\over6075} \pi^4
      +{3328\over135}\A4       -{416\over405} \pi^2 \LGdd
\right.\cr
&\left.
      +{416\over405} \LGdq     +{1561\over675}\Z3
      -{940\over243} \pi^2     +{36653\over1800}
 \right) \lr
 +U_2
      +{10681\over540}\Z5
\cr
&
      +{3328\over135}\A5
      -{416\over2025} \LGdc
      -{263\over81} \pi^2\Z3
      +{416\over1215} \pi^2 \LGdt
      -{3679\over12150} \pi^4 \LGd
\cr
&
      +{202289\over583200} \pi^4
      +{2176\over2025}\A4
      -{272\over6075} \pi^2 \LGdd
      +{272\over6075} \LGdq
      -{2933101\over60750}\Z3
\cr
&
      +{2984\over243} \pi^2 \LGd
      -{56345\over8748} \pi^2
      +{13416707\over364500}
            +O(\r^2) \ ,}  \eqno(9) $$
$$ \eqalign { \amu[{\rm fig.1(j)}]=
&{1\over8} \lrd
+\left( -{1\over2}\Z3 +{5\over12} \right) \lr
+U_3
        -{5\over6}\Z3 +{1\over48} \pi^2 +{409\over1152}
\cr
&
+\r\left[
        {110353\over272160} \pi^6
       +{9725\over756} \pi^5 \LGd
       +{4642\over567} \pi^4 \LGdd
       -{389\over8} \pi^3\Z3
       -{3376\over63} \pi^3 \Qtw
\right.\cr
&
\phantom{+\r[}
\left.
       +{1688\over189} \pi^3 \GC2\LGd
       -{6224\over63} \pi^2\GC4
       -{3376\over189} \pi^2\GCdd
       +{3376\over189} \pi^2\A4
       +{422\over567} \pi^2 \LGdq
\right.\cr
&
\phantom{+\r[}
\left.
       -{1461799\over127008} \pi^5
       -{21247\over756} \pi^4 \LGd
       -{30997\over4410} \pi^3\GC2
       +{557663\over15120} \pi^2\Z3
\right.\cr
&
\phantom{+\r[}
\left.
       +{621830879\over33339600} \pi^4
       +{23060\over189} \pi^3 \LGd
       +{18448\over189} \pi^2\GC2
       -{21563\over13230} \pi^3    -{94319\over22680} \pi^2
\right]
\cr&            +O(r^2\lrt) \ ,}  \eqno(10) $$
$$ \eqalign { \amu[{\rm fig.1(k)}]=
&\left({7\over270} \pi^4 -{32\over3}\A4   +{4\over9} \pi^2 \LGdd
       -{4\over9} \LGdq -{13\over18}\Z3  +{4\over9} \pi^2 \LGd
       +{41\over81} \pi^2
\right.\cr
&\left.
-{673\over108} \right) \lr
-{7\over135} \pi^4\Z3 +{64\over3}\A4\Z3 -{8\over9} \pi^2\Z3\LGdd
+{8\over9}\Z3\LGdq
\cr
&
+{13\over9}\Ztd  -{8\over9} \pi^2\Z3 \LGd
+{73\over12}\Z5 -{32\over3}\A5 -{49\over270} \pi^4 \LGd
+{7\over162} \pi^2\Z3
\cr
&
-{4\over27} \pi^2 \LGdt +{4\over45} \LGdc
+{97\over3240} \pi^4 -{32\over9}\A4 -{2\over27} \pi^2 \LGdd
-{4\over27} \LGdq
\cr
&
+{1985\over108}\Z3 -{59\over27} \pi^2 \LGd
+{1351\over1296} \pi^2 -{6625\over1728}
            +O(r^2\lr) \ ,}  \eqno(11) $$
$$ \eqalign { \amu[{\rm fig.2(a)}]=
&
\left(
      {8\over27} \pi^2 \LGd -{20\over81} \pi^2
       -{4\over9}\Z3 +{31\over27}
\right) \lrt
+\left(
        {11\over162} \pi^4 -{32\over9}\A4
       -{8\over27} \pi^2 \LGdd
\right.\cr
&\left.
       -{4\over27} \LGdq
       -{14\over3}\Z3 +{20\over9} \pi^2 \LGd
       -{158\over81} \pi^2  +{115\over18}
\right) \lrd
\cr
&
+\left(
        {143\over18}\Z5 -{64\over9}\A5
       +{41\over405} \pi^4 \LGd +{2\over9} \pi^2\Z3
       +{16\over81} \pi^2 \LGdt +{8\over135} \LGdc
\right.\cr
&\left.
       +{119\over810} \pi^4  -{160\over9}\A4
       -{40\over27} \pi^2 \LGdd -{20\over27} \LGdq
       -{133\over9}\Z3  +{442\over81} \pi^2 \LGd
\right.\cr
&\left.
       -{1133\over243} \pi^2 +{8719\over648}
\right) \lr
+U_4
       +{547\over36}\Z5 -{160\over9}\A5 -{1\over27} \pi^2\Z3
\cr
&
       +{40\over81} \pi^2 \LGdt +{41\over162} \pi^4 \LGd
       +{4\over27} \LGdc
       -{485\over5832} \pi^4
       -{1768\over81}\A4
       -{442\over243} \pi^2 \LGdd
\cr
&
       -{221\over243} \LGdq
       -{7199\over486}\Z3
       +{3503\over729} \pi^2 \LGd
       -{34541\over8748} \pi^2
       +{31531\over2916}
\cr
&
+\r\left[
       {101\over3072} \pi^4
\right]
            +O(r^2\lrt) \ ,}  \eqno(12) $$
$$ \eqalign { \amu[{\rm fig.2(b)}]=
&\left(
      {2\over27}  \pi^4 -{35\over6}\Z3 -{16\over9} \pi^2 \LGd
     +{62\over81} \pi^2 +{227\over54}
\right) \lrd
+\left(
      -{40\over27} \pi^2\Z3
\right.\cr
&\left.
      +{409\over1080} \pi^4
      +{104\over9}\A4
      +{35\over27} \pi^2 \LGdd
      +{13\over27} \LGdq +{1475\over162}\Z3
      -{616\over81} \pi^2 \LGd
\right.\cr
&\left.
      -{187\over729} \pi^2
      +{11891\over972}
\right) \lr
+U_5
        -{1411\over144}\Z5 +{104\over9}\A5 -{26\over27} \pi^2\Z3
\cr
&
	-{35\over81} \pi^2 \LGdt
        -{803\over3240} \pi^4 \LGd
        -{13\over135} \LGdc
	-{24679\over58320} \pi^4
      +{2284\over81}\A4
\cr
&
      +{1277\over486} \pi^2 \LGdd
      +{571\over486} \LGdq
      -{11381\over1944}\Z3 -{1858\over243} \pi^2 \LGd +{58045\over8748} \pi^2
\cr
&
      +{136247\over5832}
            +O(r^2\lr) \ ,}  \eqno(13) $$
$$ \eqalign { \amu[{\rm fig.2(c)}]=
&\left(
         {385\over18}\Z5 -{11\over9} \pi^4 \LGd +{77\over27} \pi^2\Z3
        +{139\over432} \pi^4 -{1459\over135}\Z3 +{302\over405} \pi^2
\right.\cr
&\left.
        +{3067\over3240}
\right) \lr
+U_6
        +{929\over32}\Z5 -{683\over216} \pi^2\Z3 -{319\over432} \pi^4 \LGd
\cr
&
        +{4\over27} \pi^2 \LGdd +{102541\over388800} \pi^4
        -{32\over9}\A4 -{4\over27} \LGdq -{128483\over12150}\Z3
        -{58\over27} \pi^2 \LGd
\cr
&
        +{407693\over437400} \pi^2 +{621001\over72900}
            +O(r^2) \ ,}  \eqno(14) $$
$$ \eqalign { \amu[{\rm fig.2(d)}]=
&\left(
         -{2\over3} \pi^2 \LGd +\Z3 +{5\over9} \pi^2 -{31\over12}
\right)	 \lrd
+\left(
         -{2}\Ztd +{4\over3} \pi^2\Z3 \LGd
\right.\cr
&\left.
         -{10\over9} \pi^2\Z3
         -{11\over108} \pi^4 +{16\over3}\A4 +{4\over9} \pi^2 \LGdd
         +{2\over9} \LGdq +{47\over4}\Z3
\right.\cr
&\left.
         -{55\over18} \pi^2 \LGd
	 +{97\over36} \pi^2 -{1225\over144}
\right)  \lr
         +{11\over108} \pi^4\Z3
         -{16\over3}\A4\Z3
\cr
&
         -{4\over9} \pi^2\Z3 \LGdd
         -{2\over9}\Z3 \LGdq
         -{7}\Ztd +{10\over3} \pi^2\Z3 \LGd
         -{143\over24}\Z5
\cr
&
         +{16\over3}\A5
         -{167\over54} \pi^2\Z3 -{4\over27} \pi^2 \LGdt -{41\over540} \pi^4
\LGd
         -{2\over45} \LGdc
         -{1153\over12960} \pi^4
\cr
&
         +{110\over9}\A4 +{55\over54} \pi^2 \LGdd
         +{55\over108} \LGdq
         +{461\over24}\Z3
         -{367\over108} \pi^2 \LGd
\cr
&
	 +{1871\over648} \pi^2 -{3497\over432}
\cr
&
+\r\left[
 {101\over144} \pi^3\GC2 -{101\over144} \pi^4 \LGd +{707\over288} \pi^2\Z3
+{9035\over13824} \pi^4
\right.\cr
&
\phantom{+\r[}
\left.
-{821\over864} \pi^3 -{5081\over1296} \pi^2
\right]
            +O(r^2\lrt) \ ,}  \eqno(15) $$
$$ \eqalign { \amu[{\rm fig.2(f)}]=
&
\left(
-{1\over18} \pi^4
+{35\over8}\Z3
+{4\over3} \pi^2 \LGd
-{31\over54} \pi^2
-{227\over72}
\right) \lr
+{1\over9} \pi^4\Z3
\cr
&
-{35\over4}\Ztd
-{8\over3} \pi^2\Z3 \LGd
+{46\over27} \pi^2\Z3
-{1027\over8640} \pi^4
-{13\over3}\A4
\cr
&
-{35\over72} \pi^2 \LGdd
-{13\over72} \LGdq
+{923\over864}\Z3
+{62\over27} \pi^2 \LGd
+{163\over486} \pi^2
-{4243\over1296}
\cr&            +O(r^2\lr) \ .}  \eqno(16) $$
Here $\zeta(p)$ is the Riemann $\zeta$-function of argument $p$,
\d$ \zeta(p) \equiv \sum_{n=1}^{\infty} {1\over n^p} \ ,$
$a_k$ and $\beta_k$ are the constants defined as
\d$ a_k \equiv \sum_{n=1}^{\infty} {1\over {2^n n^k}}$ and
\d$\beta_k\equiv\sum_{n=0}^\oo {(-1)^n\over (2n+1)^k}$,
whose first values are respectively
$\Z2=\pi^2/6$,
$\Z3=1.202\;056\;903...$,
$\Z4=\pi^4/90$,
$\Z5=1.036\;927\;755...$
$\Z6=\pi^6/945$,
\d$     a_4 = 0.517\;479\;061... \ $,
\d$     a_5 = 0.508\;400\;579... \ $
and
\d$ \beta_1 = \pi/4$,
\d$ \beta_2 = 0.915\;965\;594...$,
\d$ \beta_3 = \pi^3/32$,
\d$ \beta_4 = 0.988\;944\;551...$ .

$\Qtw$ is the transcendentality-three constant defined as
$$ \Qtw\equiv
 \int^{\pi/2}_0 \ln^2\left( 2\cos\left( {\theta/2}\right)\right) d\theta
    = 0.550\;279\;883\;952...\  . \eqno(17) $$

The constants $\beta_2$, $\beta_4$ and $\Qtw$ come from the
values of the polylogarithmic functions
$$ \Li_2(i)   = -{1\over 8}   \zeta(2) + i\beta_2\ , $$
$$ \Li_4(i)   = -{7\over 320} \zeta(4) + i\beta_4\  $$
and
$$ S_{12}(i)  = \int_0^i  {{\ln^2(1-t)}\over 2t} dt
              = {29\over64}\Z3 -{1\over4}\pi\beta_2
               +i\left( {1\over 32}\Z2\pi-{1\over2}\Qtw \right) \ . $$

Eq.(8)-(10) and eq.(12)-(14)
contain also six different combinations $U_i$ of several integrals
with transcendentality six
containing products of logarithms, dilogarithms and  trilogarithms
which we were unable to calculate in analytical form;
an example of such integrals is \d$\int_0^1 {dx \over {x+1}}
\Li_3(x)\Li_2(-x)$.
We have calculated them by using numerical methods to high
precision:
$$ \eqalign{
 U_1=& \phantom{+0} 1.833\;055\;809\;327... \ ,\cr
 U_2=&          \phantom{} -  8.520\;012\;033\;995... \ ,\cr
 U_3=& \phantom{+0}  0.722\;470\;399\;216... \ ,\cr
 U_4=& \phantom{+} 15.183\;055\;006\;269... \ ,\cr
 U_5=& \phantom{+} 18.086\;098\;879\;723... \ ,\cr
 U_6=& \phantom{+} 42.269\;128\;989\;300... \ .\cr
} $$

We expect that these combinations of integrals
can be expressed in analytical form
using the transcendentality-six constants,
$\zeta(6)$, $\Ztd$, $\ln^6 2$, $a_6$, $\zeta(5)\LGd$, etc..
\footnote {$^1$}{
We found that the numerical value of $U_5$
coincides with that of \d${160\over9}\Z6$ for at least
the first 28 significant figures.},
in analogy with similar integrals with transcendentality five.

The calculation of the contribution of the graph of ${\rm Fig.2(e)}$
involves many groups of integrals with transcendentality six and seven,
and we were unable to complete it in analytical form without
the knowledge of the explicit analytical value of these integrals.
The contribution of this graph will be given later
in numerical form.

The expansions in \d$(m_e/m_\mu)$ eqs.(1)-(16)
present some features already found in the expressions
of the three-loop and four-loop contributions
to the muon $g$-$2$ [1][2]:

1) The expression of the contributions of the graphs
containing only electron loops
(eq.(1), (5), (10), (12) and (15))
have a linear $(m_e/m_\mu)$ term
\footnote {$^2$}{
At present
only the analytical expression
of the contribution of the three-loop light-light graphs
it is known to contain a term proportional to
$(m_e/m_\mu)$ $\ln(m_\mu/m_e)$.}.
The coefficients of these terms are constant,
being numerically respectively -4.74, -18.00, -20.38, 3.20, -2.81.
As we expected, these numerical values are larger
than those found in four-loop calculations [2].
The $(m_e/m_\mu)$ expansions of the contributions of the other
graphs
containing also muon loops begin with the $(m_e/m_\mu)^2$ term.

2) The coefficients of the expansions
   in $(m_e/m_\mu)$
   of the contributions of the
   graphs containing only electron loops
   contain high powers of $\ln(m_\mu/m_e)$.
   As an example, in table 1 we have listed
   the numerical values
   and the maximum power of $\ln(m_\mu/m_e)$
   of the coefficients of the powers
   of \d$(m_e/m_\mu)$ of eq.(5).
   We found that the coefficients of the even powers
   have the maximum power of $\ln(m_\mu/m_e)$ equal or greater than
   the maximum power of the zero order coefficient;
   the coefficient of $(m_e/m_\mu)^4$
   contains the highest power of $\ln(m_\mu/m_e)$.
   The coefficients of the odd powers (greater than one)
   contain at most a $\ln(m_\mu/m_e)$ term.
   The same behaviour of the coefficients of the expansions was found
   even in all similar sixth- and eighth-order contributions [1][2];
   this observation could be useful in the analytical calculations
   of the contributions
   of other high-order graphs containing only electron loops
   in order to estimate the magnitude of the uncalculated
   terms of the expansions in $(m_e/m_\mu)$.

3) At the five-loop level
   new transcendental constants, $\beta_4$ and $\Qtw$,
   appear  in the
   coefficient of the $(m_e/m_\mu)$ term of the
   expression of the
   contribution of the graph of Fig.1(j),
   containing a double insertion of
   a fourth-order vacuum polarization on a second order vertex graph;
   in analogous way,
   the odd power $\pi^3$
   and
   the constant $\beta_2$ appeared
   at the three-loop level [1] and four-loop level [2] respectively.
   This fact remarks that
   the not leading terms of the contribution
   of the graph of Fig.1(j) cannot be worked out
   from the not leading terms of simpler graphs,
   whereas the leading ones can be easily deduced
   using the renormalization group equations [3].

We have compared our results with
the renormalization-group results of
ref.[3].
The leading terms of our eq.(1) and (5)
agree respectively with eq.(35) and (32) of ref.[3]
The logarithmic and constant terms of our eq.(10)
agree with eq.(25) of ref.[3],
except that our expression contains
the unknown transcendentality-six constant $U_3$, whereas
the expression of ref.[3] contains the analytical term $\Ztd/2$;
these terms are in perfect numerical agreement,
so that the analytical expression of $U_3$ can be inferred.

In table 2 we have listed the numerical values of eqs.(1)-(16)
obtained
using the experimental value [4] $ {(m_\mu/m_e)}=206.768 262(30)$
and taking into account the (not shown) terms up to
$(m_e/m_\mu)^4$.
The value of the contribution of the graph of Fig.2(e)
has been worked out evaluating numerically
 the one-dimensional integral obtained
using the dispersive representation for the vacuum polarization [2].

As expected, the graphs containing muon loops give contributions
substantially
smaller than those containing only electron loops.
The sum of the numerical values of the three graphs of fig.1
calculated using the renormalization group technique
in ref.[3] has a rather large value, about 52.
On the contrary the sum of all contributions of fig.1 and 2
shows a strong cancellation,
the numerical value being
$$ \amu[\fig{1+2}] =-1.261\;574(2) \ . \eqno(18)$$
This fact is due to the cancellation between
the positive contributions of the graphs of fig.1
containing vacuum polarization
insertion on the second-order vertex graph
and the negative contributions of
the graphs of fig.2 containing vacuum
polarization insertions on fourth-order vertex graphs.
We found a similar behaviour at the four-loop level,
even if in that case the cancellation was less marked [2].
Finally, eq.(18) turns out to be much smaller than the estimate
of the total tenth-order contribution [5]
$$ \amu            = 570(140) \ .  \eqno(19) $$

\par
\vskip 6 truemm
All the algebraic manipulations were carried out through the symbolic
manipulation program ASHMEDAI [6].
\par
\vfill\eject
{\bf References}
\vskip 10 truemm

\item{[1]}  S.Laporta,  {\it Nuovo Cimento   A},{\bf 106}, (1993) 675.
\item{[2]}  S.Laporta,  {\it Phys.Lett.} {\bf B}312, (1993) 495.
\item{[3]}  A.L.Kataev, {\it Phys.Lett.} {\bf B}284, (1992) 401.
\item{[4]} E. R. Cohen and B. N. Taylor, {\it Rev. Mod. Phys.},
           {\bf 59}, (1987) 1121.
\item{[5]}  T.Kinoshita and W.B.Lindquist, \PR D {\bf 41}, (1990) 593;
\par
            T.Kinoshita and W.Marciano in {\it Quantum Electrodynamics},
            edited by T.Kinoshita \par
            (World Scientific, Singapore, 1990), p.419.
\parn
\item{[6]} M. J. Levine, U.S. AEC Report No. CAR-882-25 (1971)
                                                       (unpublished).
\parn
\vfill\eject
{\bf Figure captions}
\vskip 12 truemm
\parn
Fig.1: Tenth-order vertex graph
       obtained with insertions of second- and fourth-order
       vacuum polarization subdiagrams on the second-order vertex graph.
\parn
Fig.2: Examples of tenth-order vertex graph
       obtained with insertions of second- and fourth-order
       vacuum polarization subdiagrams on the fourth-order vertex graphs.
\parn
\vskip 12 truemm
\vfill\eject
\riga
TABLE 1: The numerical values and the maximum power of $\ln(m_\mu/m_e)$
         of the coefficients of the expansion
 in $\left({m_e/m_\mu}\right)$ of the contribution of the
graph of Fig.1(e).
{
\riga
\hskip 9 truemm
 {\it n}  \hskip 5truemm
 Coefficient of
    \hskip 10 truemm
 Maximum power
\parn
\hskip 9 truemm
\phantom {n}
\hskip 8truemm
 $(m_e/m_\mu)^n$
\hskip 15 truemm
 of $\ln(m_\mu/m_e)$
\riga
$$
 \eqalign {
&0 \qquad \cr
&1 \cr
&2 \cr
&3 \cr
&4 \cr
&5 \cr
&6 \cr
&7 \cr
&8 \cr
&9 \cr
&10 \cr
}
 \eqalign {
&27.71         \hskip 30 truemm  {} \cr
-&18.00   \cr
&2493.30   \cr
&154.49   \cr
-&34783.30 \cr
-&3.91    \cr
&452.06   \cr
&11.93    \cr
-&3859.59  \cr
&3.15     \cr
-&1346.48  \cr
}
 \eqalign {
&3 \hskip 86 truemm  {} \cr
&0 \cr
&4 \cr
&0 \cr
&5 \cr
&0 \cr
&4 \cr
&0 \cr
&4 \cr
&0 \cr
&4 \cr
}
 $$
\riga
}

\vfill\eject
\riga
TABLE 2: The numerical values of the contributions to $\amu$
of the graphs of figs.1 and 2.
\riga
\hskip 7 truemm
 Figure  \hskip 14truemm
 Contribution to $\amu$
\riga
$$ \eqalign{
&{\rm 1(a)}       \hskip 19 truemm  {}\cr
&{\rm 1(b)} \cr
&{\rm 1(c)} \cr
&{\rm 1(d)} \cr
&{\rm 1(e)} \cr
&{\rm 1(f)} \cr
&{\rm 1(g)} \cr
&{\rm 1(h)} \cr
&{\rm 1(i)} \cr
&{\rm 1(j)} \cr
&{\rm 1(k)} \cr
&{\rm 2(a)} \cr
&{\rm 2(b)} \cr
&{\rm 2(c)} \cr
&{\rm 2(d)} \cr
&{\rm 2(e)} \cr
&{\rm 2(f)} \cr
&{\rm 1+2} \cr
}
 \eqalign{
& 20.142\;811   (3)         \hskip 92 truemm  {}  \cr
&  2.203\;327\;2 (2)      \cr
&  0.206\;959\;08(1)      \cr
&  0.013\;875\;908\;8(3) \cr
& 27.690\;059\;(3)       \cr
&  1.166\;152\;15(6)     \cr
&  0.031\;814\;813\;1(6) \cr
&  1.614\;350\;1(1)      \cr
&  0.164\;714\;747(4)    \cr
&  4.742\;149\;1 (2)     \cr
&  0.399\;245\;484 (8)   \cr
& -28.429\;744\;(2)      \cr
&  -6.792\;291\;9(3)     \cr
&  -0.952\;823\;49 (2)   \cr
& -19.042\;323\;(1)      \cr
&  -3.131\;386\;2(1)     \cr
&  -1.288\;464\;12(2)    \cr
& -1.261\;574(2)
} $$
\riga
\bye